\documentclass[a4paper,fleqn]{cas-dc}

\usepackage[numbers]{natbib}

\usepackage{bm}

\begin{document}
\let\WriteBookmarks\relax
\def\floatpagepagefraction{1}
\def\textpagefraction{.001}
\shorttitle{}
\shortauthors{}

\title [mode = title]{First-forbidden transition of nuclear $\beta$ decay by projected shell model}




\author[1]{Bin-Lei Wang}
\address[1]{School of Physical Science and Technology, Southwest University, Chongqing 400715, China}

\author[1]{Long-Jun Wang}
\cormark[1]

\cortext[cor1]{longjun@swu.edu.cn}

\begin{abstract}
  The first-forbidden transition of nuclear $\beta$ decay is expected to play crucial roles in many aspects in nuclear physics, nuclear astrophysics and particle physics such as the stellar $\beta$-decay rates and the reactor anti-neutrino spectra. In this work we develop the projected shell model (PSM) for description of first-forbidden transition of nuclear $\beta$ decay for the first time. Detailed theoretical framework and logics are provided, and 35 dominant first-forbidden transitions that are expected to be important for the reactor anti-neutrino spectra problems are calculated and compared systematically with the data to test the new development of the PSM. The corresponding experimental Log$f_0 t$ values are described reasonably, and the quenching factors of nuclear matrix elements are found to affect the Log$f_0 t$ values as well as the related shape factors, which may be helpful for better understanding of the reactor anti-neutrino spectra problems. 
\end{abstract}




\maketitle

Nuclear $\beta$ decay plays important and indispensable roles in many aspects in nuclear physics, nuclear astrophysics, particle physics and nuclear engineering. The understanding of the origin of heavy elements rely heavily on stellar $\beta$-decay and electron-capture rates as inputs for studying the core-collapse supernovae, the slow ($s$-) and rapid ($r$-) neutron-capture process etc. \cite{Fuller1980, Fuller1982_1, Fuller1982_2, fuller1985, rp_process_Schatz_1998, langanke_RMP, r_process_RMP_2021, langanke_2021_Rep_Pro_Phys}, while the $\beta$ spectrum and anti-neutrino spectrum are important for the up-down quark element of the Cabibbo-Kobayashi-Maskawa matrix, the reactor anti-neutrino anomaly, the calculations of residual power or post-irradiation fuel management etc. \cite{Hayen_RMP_2018, Hayen_CPC_code_2019, Fallot_PRL_2012, An_PRL_2016, An_PRL_2022, Almaz_PRL_2022}

Two kinds of transitions are involved in nuclear $\beta$ decay: allowed and forbidden, according to the angular momentum ($l$) taken by the emitted leptons relative to the nucleus. Allowed transition corresponds to $l=0$ while $l>0$ ($l=1$) corresponds to the (first) forbidden transition. The properties of nuclear $\beta$ decays are dominated by allowed transition in most cases while it is the (first) forbidden transition that plays crucial and decisive roles in many cases. For stellar $\beta$ decays, the first-forbidden transitions are expected to reduce the half-lives of the important $N=126$ $r$-process waiting-point nuclei \cite{Zhi_FF_PRC_2013}, and the stellar $\beta$-decay rates of many important $s$-process branching-point nuclei, such as $^{151}$Sm, $^{170}$Tm, $^{185}$W, $^{204}$Tl etc. are determined by first-forbidden transitions since nearly all of the relevant transitions are first-forbidden \cite{TY_table_1987}. For the reactor anti-neutrino spectra problems, such as the reactor anti-neutrino anomaly that corresponds to a $6\%$ deficit between experimental anti-neutrino count rate and theoretical predictions, and the reactor anti-neutrino shoulder that corresponds to a spectral disagreement (bump) between observations against theoretical predictions at $4-6$ MeV of anti-neutrino energy, the dominant first-forbidden transitions are expected to mitigate and for better understanding the problems \cite{Fallot_PRL_2012, Ahn_PRL_2012, Abe_JHep_2014, An_PRL_2016, An_PRL_2022, Almaz_PRL_2022, Hayen_PRC_R_2019, Hayen_PRC_2019, 137Xe_FF_PRL_2020}. 

In the above discussions, on one hand, for stellar $\beta$-decay rates, (parent) nuclei have probability to be thermally populated in excited states in the stellar environments with high temperature and density, transitions from excited states are then indispensable which are however beyond the current experimental ability. On the other hand, for the reactor anti-neutrino spectra problems, numerous fission fragments are exotic nuclei with $A=85-145$ so that only parts of the relevant spectra have been measured and with poor precision \cite{137Xe_FF_PRL_2020}. Theoretical description of first-forbidden transitions are then desired, which is not as straightforward as the allowed transition. Compared with the latter, the transition operators and the corresponding phase space for first-forbidden transitions become much complicated \cite{Weidenmuller_FF_RMP_1961, Daniel_RMP_1968, Suhonen_book}, and change of parity is further embedded which calls for large model space to include orbitals with opposite parity. Over the decades, many nuclear many-body methods have been developed to describe allowed and (first) forbidden transitions of nuclear $\beta$ decays. Those methods are mainly based on the conventional shell model (SM) or the quasiparticle random phase approximation (QRPA) \cite{Zhi_FF_PRC_2013, Kumar_PRC_2020, Marketin_QRPA_FF_PRC_2016, Sharma_PRC_2022, shell_model_RMP_2005, HKWang_2022_PLB, HKWang_2022_PRC, Suhonen_book, P.sarriguren2001NPA, Engel_1999_PRC, YFNiu_2015_PRL}. 

Recently, the projected shell model (PSM) with large model space \cite{PSM_review, Sun_1996_Phys_Rep} is extended to have high-order quasiparticle (qp) configurations with large configuration space, with the help of the Pfaffian and other algorithms \cite{Mizusaki_2013_PLB, LJWang_2014_PRC_Rapid, LJWang_2016_PRC, ZRChen_2022_PRC, BLWang_2022_PRC}. The PSM is then applied to studies on nuclear high-spin physics \cite{LJWang_2016_PRC, LJWang_2019_JPG, LJWang_PLB_2020_chaos}. Besides, the PSM is further developed for describing allowed Gamow-Teller (GT) transition and stellar weak-interaction rates \cite{Z_C_Gao_2006_GT, LJWang_2018_PRC_GT, LJWang_PLB_2020_ec, LJWang_2021_PRL, LJWang_2021_PRC_93Nb, zrchen2023symm, ZRChen_PLB2023} and nuclear $\beta$ spectrum \cite{FGao_PRC2023}. In this work we present a new development of the extended PSM for description of first-forbidden transition of nuclear $\beta$ decay for the first time. This new development should be meaningful as the PSM is more successful for heavier nuclei where first forbidden transition tends to become the major decay mode in nuclear $\beta$ decay.


For a first-forbidden transition from an initial state (usually ground state or isomers, with spin-parity $J_i^{\pi_i}$) to a final state ($J_f^{\pi_f}$), the corresponding partial half-life, $t$, can be obtained by \cite{Zhi_FF_PRC_2013}, 
\begin{eqnarray} \label{eq.ft}
  ft = K_0 = 6146 \text{ s},
\end{eqnarray}
where the phase-space integral for $\beta^-$ decay reads as,
\begin{eqnarray} \label{eq.phase_f}
  f = \int_{1}^{Q_{if}} C(W) F_0(Z, W) \ W p \  (Q_{if} - W)^{2} dW ,
\end{eqnarray}
with $W$ and $p=\sqrt{W^{2}-1}$ being the total energy (rest mass and kinetic energy) and the momentum of the electron in units of $m_{e}c^{2}$ and $m_{e}c$, respectively. The available total energy for leptons is given by,
\begin{eqnarray} \label{Qif}
  Q_{if} = \frac{1}{m_e c^2} ( \Delta M_{pd} + E_i - E_f),
\end{eqnarray}
where $\Delta M_{pd}$ denotes the nuclear mass difference of the parent and daughter nuclei, $E_i$ and $E_f$ are the excitation energies of initial ($i$) and final ($f$) states. 

In Eq. (\ref{eq.phase_f}), $F_0(Z, W)$ is the Fermi function that reveals the Coulomb distortion of the electron wave function near the nucleus \cite{Fuller1980, Fermi_func_1983}, and the important shape factor can be approximated as \cite{Weidenmuller_FF_RMP_1961, Zhi_FF_PRC_2013, Mougeot_PRC_2015}
\begin{eqnarray} \label{eq.CW}
  C(W) = k(1 + aW + b/W + cW^2) ,
\end{eqnarray}
for non-unique transitions with $|\Delta J| = |J_i - J_f| = 0,1$, $\Delta \pi=\pi_i \pi_f=-1$, and \cite{Mougeot_PRC_2015}
\begin{eqnarray} \label{eq.CW2}
  C(W) = k(1 + aW + b/W + cW^2) \left[ q^2 + \lambda_2 p^2 \right] ,
\end{eqnarray}
for unique transitions with $|\Delta J|=2$, $\Delta\pi=-1$, where $q=Q_{if} - W$ is the energy of the neutrino and $\lambda_2 = \frac{F_1(Z, W)}{F_0(Z, W)}$ with $F_1$ being the generalized Fermi function \cite{Suhonen_PRC_2017_general_Fermi_for_unique}. The $\beta$ spectrum, i.e., the integrand in Eq. (\ref{eq.phase_f}), and the corresponding anti-neutrino spectrum, are related to the shape factor where the coefficients $k, a, b$ and $c$ depend on reduced nuclear matrix elements of first-forbidden transition as follows \cite{Behrens_NPA_1971}, 
\begin{eqnarray} \label{eq.kabc}
  k  &=& [\zeta_0^2 + \frac{1}{9} w^2]^{(0)} + [\zeta_1^2 + \frac{1}{9}(x+u)^2 - \frac{4}{9}\mu_1\gamma_1 u(x+u)   \nonumber \\
     & & \quad + \frac{1}{18} Q_{if}^2 (2x+u)^2 - \frac{1}{18} \lambda_2 (2x-u)^2 ]^{(1)} \nonumber \\
     & & \quad + [\frac{1}{12} z^2 (Q_{if}^2-\lambda_2)]^{(2)} , \nonumber \\
  ka &=& [-\frac{4}{3} u Y -\frac{1}{9} Q_{if} (4x^2 +5u^2)]^{(1)} -[\frac{1}{6} z^2 Q_{if}]^{(2)} , \nonumber \\
  kb &=& \frac{2}{3} \mu_1 \gamma_1 \{ -[\zeta_0 w]^{(0)} + [\zeta_1 (x+u)]^{(1)} \} , \nonumber \\
  kc &=& \frac{1}{18} [8u^2 +(2x+u)^2 + \lambda_2 (2x-u)^2]^{(1)} \nonumber \\
     & & \quad +\frac{1}{12} [z^2 (1+\lambda_2)]^{(2)} ,
\end{eqnarray}
where the numbers in parentheses of superscripts denote the rank of the transition operators, $\xi = \alpha Z / (2R)$ with $R$ being the radius of the nucleus and $\alpha$ the fine structure constant, the parameter $\gamma_1=\sqrt{1 - (\alpha Z)^2}$. For the Coulomb functions $\mu_1$ and $\lambda_2$ we adopt the approximations $\mu_1 \approx 1$, $\lambda_2 \approx 1$ as in Refs. \cite{Zhi_FF_PRC_2013, Mougeot_PRC_2015, Marketin_QRPA_FF_PRC_2016}. The $\zeta_0, \zeta_1$ and $Y$ in Eq. (\ref{eq.kabc}) are defined as, 
\begin{align} \label{eq.VY}
  V &= \xi' \nu + \xi w',   \qquad \ \quad \zeta_0 = V + \frac{1}{3} w Q_{if}, \nonumber \\
  Y &= \xi'y - \xi (u'+x'), \quad  \zeta_1 = Y + \frac{1}{3} (u-x) Q_{if} .  
\end{align}

After a non-relativistic reduction and adopting the Condon-Shortley phase convention, the reduced nuclear matrix elements shown in Eqs. (\ref{eq.kabc}, \ref{eq.VY}) are \cite{Behrens_NPA_1971, Zhi_FF_PRC_2013}, 
\begin{subequations} \label{eq.all_ME}
\begin{eqnarray}
  w  &=& -g_A \sqrt{3} \frac{\left\langle \Psi^{n_f}_{J_f} \left\| \sum_k r_k [\bm C^k_1 \otimes \hat{\bm\sigma}^k]^0 \hat{\tau}^k_- \right\| \Psi^{n_i}_{J_i} \right\rangle}{\sqrt{2J_i+1}} , \\
  x  &=& - \frac{\left\langle \Psi^{n_f}_{J_f} \left \| \sum_k r_k \bm C^k_1 \hat{\tau}^k_- \right \| \Psi^{n_i}_{J_i} \right \rangle}{\sqrt{2J_i+1}} , \\
  u  &=& -g_A \sqrt{2} \frac{\left\langle \Psi^{n_f}_{J_f} \left \| \sum_k r_k [\bm C^k_1 \otimes \hat{\bm\sigma}^k]^1 \hat\tau^k_- \right \| \Psi^{n_i}_{J_i} \right \rangle}{\sqrt{2J_i+1}} , \\
  z  &=& 2g_A \frac{\left\langle \Psi^{n_f}_{J_f} \left \| \sum_k r_k [\bm C^k_1 \otimes \hat{\bm\sigma}^k]^2 \hat\tau^k_- \right \| \Psi^{n_i}_{J_i} \right \rangle}{\sqrt{2J_i+1}} , \\ 
  w' &=& -g_A \sqrt{3} \frac{\left\langle \Psi^{n_f}_{J_f} \left \| \sum_k \frac{2}{3}r_k I(r_k) [\bm C^k_1 \otimes \hat{\bm\sigma}^k]^0 \hat\tau^k_- \right \| \Psi^{n_i}_{J_i} \right \rangle}{\sqrt{2J_i+1}} , \nonumber \\ \\
  x' &=& - \frac{\left\langle \Psi^{n_f}_{J_f} \left \| \sum_k \frac{2}{3}r_k I(r_k)  \bm C^k_1 \hat\tau^k_- \right \| \Psi^{n_i}_{J_i} \right \rangle}{\sqrt{2J_i+1}} , \\
  u' &=& -g_A \sqrt{2} \frac{\left\langle \Psi^{n_f}_{J_f} \left \| \sum_k  \frac{2}{3}r_k I(r_k) [\bm C^k_1 \otimes \hat{\bm\sigma}^k]^1 \hat\tau^k_- \right \| \Psi^{n_i}_{J_i} \right \rangle}{\sqrt{2J_i+1}} , \nonumber \\ \\
  \xi'\nu &=& \frac{g_A\sqrt{3}}{M_0} \frac{\left\langle \Psi^{n_f}_{J_f} \left \| \sum_k  [\hat{\bm\sigma}_k \otimes \bm\nabla^k]^0 \hat\tau^k_- \right \| \Psi^{n_i}_{J_i} \right \rangle}{\sqrt{2J_i+1}} , \\
  \xi' y  &=& - \frac{1}{M_0} \frac{\left\langle \Psi^{n_f}_{J_f} \big\| \sum_k  \bm\nabla^k \hat\tau^k_- \big\| \Psi^{n_i}_{J_i} \right \rangle}{\sqrt{2J_i+1}}  .
\end{eqnarray} 
\end{subequations}
where $\hat{\bm \sigma}$ is the Pauli spin operator, $\hat\tau_-$ is the isospin lowering operator with the convention $\hat\tau_- |n\rangle = |p\rangle$. $\Psi^n_J$ represents the nuclear many-body wave function of the $n$-th eigen-state for angular momentum $J$, $M_0$ labels the nucleon mass, and the weak axial coupling constant is $g_A = -1.2701(25)$. 
\begin{eqnarray}
   \bm C_{lm} = \sqrt{ \frac{4\pi}{2l+1} } \bm Y_{lm}, 
\end{eqnarray}
with $\bm Y_{lm}$ being the spherical harmonics. The radial function $I(r)$ takes into account the nuclear charge distribution which is usually approximated by a uniform spherical distribution, so that,
\begin{eqnarray}
  I(r) = \frac{3}{2}
  \left\{ \begin{array}{cl} 
      \left[1-\frac{1}{5}(\frac{r}{R})^2\right]              & 0 \leqslant r \leqslant R, \\ 
      \left[\frac{R}{r} - \frac{1}{5}(\frac{R}{r})^3 \right] & r \geqslant R. 
  \end{array} \right.  
\end{eqnarray}

To compare with experimental measurements, the comparative half-life, $f_0 t$, is usually needed with $f_0$ being
\begin{eqnarray} \label{eq.phase_f0}
  f_0 = \int_{1}^{Q_{if}} F_0(Z, W)  W p  (Q_{if} - W)^{2} dW 
\end{eqnarray}
for non-unique transitions and
\begin{eqnarray} \label{eq.phase_f02}
  f_0 = \int_{1}^{Q_{if}} F_0(Z, W)  W p  (Q_{if} - W)^{2} \left[ q^2 + \lambda_2 p^2 \right] dW 
\end{eqnarray}
for unique transitions. 

As the first-forbidden transitions have strong selection rules, the nuclear many-body wave function $\Psi^n_J$ should be written in the laboratory system with good angular momentum and parity. In the PSM, this is accomplished by the angular-momentum projection (AMP) techniques, i.e. \cite{PSM_review}, 
\begin{eqnarray} \label{eq.wave_function}
  | \Psi^{n}_{JM} \rangle = \sum_{K\kappa} f_{K\kappa}^{Jn} \hat{P}_{MK}^{J} | \Phi_{\kappa} \rangle ,
\end{eqnarray}
where $f$ labels the expansion coefficients in the projected basis with $\hat{P}_{MK}^{J}$ being the AMP operator as,
\begin{eqnarray} \label{AMP_operator}
    \hat{P}^{J}_{MK} = \frac{2J + 1}{8\pi^2} \int d\Omega D^{J\ast}_{MK} (\Omega) \hat{R} (\Omega) ,
\end{eqnarray}
where $\hat{R}$ and $D_{MK}^{J}$ (with Euler angle $\Omega$) \cite{varshalovich1988quantum} are the rotation operator and Wigner $D$-function \cite{BLWang_2022_PRC} respectively, $K$ ($M$) is the AMP in the intrinsic (laboratory) system. In Eq. (\ref{eq.wave_function}) $|\Phi_{\kappa}\rangle$ is the many-body multi-qp configurations which read as
\begin{align} \label{eq.config}
  \textrm{ee}: \big\{ & |\Phi \rangle, 
               \hat{a}^\dag_{\nu_i} \hat{a}^\dag_{\nu_j} |\Phi \rangle,
               \hat{a}^\dag_{\pi_i} \hat{a}^\dag_{\pi_j} |\Phi \rangle,
               \hat{a}^\dag_{\nu_i} \hat{a}^\dag_{\nu_j} \hat{a}^\dag_{\pi_k} \hat{a}^\dag_{\pi_l} |\Phi \rangle, \nonumber\\ 
             & \hat{a}^\dag_{\nu_i} \hat{a}^\dag_{\nu_j} \hat{a}^\dag_{\nu_k} \hat{a}^\dag_{\nu_l} |\Phi \rangle, 
               \hat{a}^\dag_{\pi_i} \hat{a}^\dag_{\pi_j} \hat{a}^\dag_{\pi_k} \hat{a}^\dag_{\pi_l} |\Phi \rangle,  \cdots \big\} \nonumber \\
  \textrm{o}\nu: \big\{ & \hat{a}^\dag_{\nu_i} |\Phi \rangle,
                 \hat{a}^\dag_{\nu_i} \hat{a}^\dag_{\nu_j} \hat{a}^\dag_{\nu_k} |\Phi \rangle, 
                 \hat{a}^\dag_{\nu_i} \hat{a}^\dag_{\pi_j} \hat{a}^\dag_{\pi_k} |\Phi \rangle,   \nonumber \\
               & \hat{a}^\dag_{\nu_i} \hat{a}^\dag_{\nu_j} \hat{a}^\dag_{\nu_k} \hat{a}^\dag_{\pi_l} \hat{a}^\dag_{\pi_m} |\Phi \rangle,
                 \cdots \big\} \nonumber \\
  \textrm{o}\pi: \big\{ & \hat{a}^\dag_{\pi_i} |\Phi \rangle,
                 \hat{a}^\dag_{\pi_i} \hat{a}^\dag_{\pi_j} \hat{a}^\dag_{\pi_k} |\Phi \rangle, 
                 \hat{a}^\dag_{\pi_i} \hat{a}^\dag_{\nu_j} \hat{a}^\dag_{\nu_k} |\Phi \rangle,   \nonumber \\
               & \hat{a}^\dag_{\pi_i} \hat{a}^\dag_{\pi_j} \hat{a}^\dag_{\pi_k} \hat{a}^\dag_{\nu_l} \hat{a}^\dag_{\nu_m} |\Phi \rangle,
                 \cdots \big\} \nonumber \\
  \textrm{oo}: \big\{ & \hat{a}^\dag_{\nu_i} \hat{a}^\dag_{\pi_j}|\Phi \rangle, 
               \hat{a}^\dag_{\nu_i} \hat{a}^\dag_{\nu_j} \hat{a}^\dag_{\nu_k} \hat{a}^\dag_{\pi_l} |\Phi \rangle,
               \hat{a}^\dag_{\nu_i} \hat{a}^\dag_{\pi_j} \hat{a}^\dag_{\pi_k} \hat{a}^\dag_{\pi_l} |\Phi \rangle, \nonumber\\ 
             & \hat{a}^\dag_{\nu_i} \hat{a}^\dag_{\nu_j} \hat{a}^\dag_{\nu_k}  
               \hat{a}^\dag_{\pi_l} \hat{a}^\dag_{\pi_m} \hat{a}^\dag_{\pi_n} |\Phi \rangle,  \cdots \big\} 
\end{align}
for even-even (ee), odd-neutron (o$\nu$), odd-proton (o$\pi$) and odd-odd (oo) nuclei respectively, where $|\Phi \rangle$ is the qp vacuum with associated intrinsic deformation and $\hat{a}^\dag_\nu (\hat{a}^\dag_\pi)$ labels neutron (proton) qp creation operator. The large configuration spaces in Eq. (\ref{eq.config}) are developed in Refs. \cite{LJWang_2014_PRC_Rapid, LJWang_2016_PRC, LJWang_2018_PRC_GT} recently and three or more major harmonic-oscillator shells can be adopted for model space. 

The kernels for first forbidden transitions, i.e., the different reduced nuclear matrix elements in Eq. (\ref{eq.all_ME}) can be calculated by,
\begin{align} \label{eq.RME}
   & \left\langle \Psi^{n_f}_{J_f} \left\| \hat{\mathcal O}_\lambda \hat\tau_{-} \right\| \Psi^{n_i}_{J_i} \right\rangle \nonumber \\
  =& \hat\lambda^{-1} \sum_{\pi_i \nu_j} \langle \pi_i \| \hat{\mathcal O}_\lambda \hat\tau_{-} \| \nu_j \rangle 
     \left\langle \Psi^{n_f}_{J_f} \left\| \left[ \hat c^\dag_{\pi_i} \otimes \tilde{\hat c}_{\nu_j} \right]^\lambda \right\| \Psi^{n_i}_{J_i} \right\rangle ,
\end{align}
where $\hat\lambda \equiv \sqrt{2\lambda+1}$ with $\lambda$ being the rank of the operators $\hat{\mathcal O}_\lambda$ in Eq. (\ref{eq.all_ME}), $\pi_i$ ($\nu_j$) labels single-particle proton (neutron) state for $\beta^-$ decay. Here we adopt the spherical harmonic oscillator basis so that $|\mu\rangle \equiv |n_\mu, l_\mu, j_\mu \rangle$ and $\hat c^\dag, \tilde{\hat c}$ are the corresponding single-particle creation and annihilation operators in the form of irreducible spherical tensor. The last term in Eq. (\ref{eq.RME}) is the reduced one-body transition density, which is derived in the AMP framework as \cite{FGao_PRC2023},
\begin{align} \label{eq.ROBTD}
  & \big\langle \Psi^{n_f}_{J_f} \big\| \left[ \hat c^\dag_\mu \otimes \tilde{\hat c}_\nu \right]^\lambda \big\| \Psi^{n_i}_{J_i} \big\rangle 
  \nonumber \\
  =&\ \sqrt{2J_f+1}
      \sum_{K\kappa K'\kappa'} f^{J_f n_f}_{K \kappa} f^{J_i n_i}_{K' \kappa'} 
      \sum_\rho C^{J_f K}_{J_i K-\rho \lambda \rho} \nonumber \\
      & \quad \times
      \frac{2J_i+1}{8\pi^2} \int d\Omega D^{J_i \ast}_{K-\rho K'}(\Omega)
      \sum_{m_\mu m_\nu} C_{j_\mu m_\mu j_\nu m_\nu}^{\lambda \rho}  \nonumber \\
      & \quad \times
      (-)^{j_\nu + m_\nu}
      \big\langle \Phi_{\kappa} \big|  
      \hat c^\dag_{j_\mu m_\mu}  {\hat c}_{j_\nu -m_\nu} 
      \hat R(\Omega) \big| \Phi_{\kappa'} \big\rangle  .
\end{align}
where $C$ labels the Clebsch-Gordan coefficients, and the Condon-Shortley phase convention is adopted in all cases here. The last term in Eq. (\ref{eq.ROBTD}) is the rotated matrix element that can be calculated by the Pfaffian algorithm in the PSM framework as shown in Ref. \cite{FGao_PRC2023} in details. 

As in the allowed GT transition of $\beta$ decay \cite{A.brown1985, martinez1996} and in the double $\beta$ decay \cite{Javier2011PRL, LJWang_current_2018_Rapid}, quenching factors for nuclear matrix elements of first-forbidden transitions of $\beta$ decay should probably be introduced as well \cite{Zhi_FF_PRC_2013, Marketin_QRPA_FF_PRC_2016, Sharma_PRC_2022}. In the SM studies of Ref. \cite{Zhi_FF_PRC_2013} the following quenching factors for the various matrix elements in Eq. (\ref{eq.all_ME}) are adopted,
\begin{align} \label{eq.quench}
  f_q(\xi'\nu) &= 1.266,          \hspace{3.95em} f_q(w) = f_q(w') = 0.66, \nonumber \\
  f_q(x)       &= f_q(x') = 0.51, \quad f_q(u) = f_q(u') = 0.38, \nonumber \\
  f_q(z)       &= 0.42. 
\end{align}




\begin{table*}[width=2.1\linewidth,cols=16,pos=t]
  \caption{Dominant first-forbidden transitions for the reactor anti-neutrino spectral shoulder \cite{Hayen_PRC_R_2019, Hayen_PRC_2019}. Here $Q_\beta$ is the ground-state to ground-state $Q$ value in MeV \cite{AME2020}, $E_f$ is the excitation energy of the final states in MeV. The PSM calculations for the Log$f_0 t$ and the coefficients $k, a, b, c$ in the shape factor of Eqs. (\ref{eq.CW}, \ref{eq.CW2}) without ($f_q=1$) and with the SM ($f_q$ SM) quenching factors in Eq. (\ref{eq.quench}) are compared with the data \cite{NNDC, livechart} when available. } \label{tab1}
\begin{tabular*}{\tblwidth}{@{} LCCCCCCCCCCCCCCC@{} }
  \toprule
  Nuclei & $Q_\beta$ & $E_f$ & $J_i^\pi \rightarrow J_f^\pi$ & $|\Delta J|$ & Log$f_0 t$ 
  & \multicolumn{5}{c}{$f_q$=1} & \multicolumn{5}{c}{$f_q$ SM} \\ \cline{7-11} \cline{12-16}
         &     &   &  &  & (Exp) & Log$f_0 t$ & $k$ & $a$ & $b$ & $c$ & Log$f_0 t$ & $k$ & $a$ & $b$ & $c$ \\
  \midrule
  $^{89}$Br  & 8.26 & 0.03 & $5/2^- \rightarrow 5/2^+$ & 0 & 6.5   
             & 6.96 & 5.58E-4 & -6.35E-2 & 4.35E-2 & 8.99E-3 & 6.57 & 1.65E-3 & -3.93E-3 & 2.83E-2 & 4.63E-4 \\
  $^{90}$Rb  & 6.58 & 0    & $0^- \rightarrow 0^+$     & 0 & 7.35 
             & 9.04 & 5.51E-6 & 0 & 1.07E-1 & 0 & 8.26 & 3.38E-5 & 0 & 2.86E-2 & 0 \\
  $^{91}$Kr  & 6.77 & 0.11 & $5/2^+ \rightarrow 5/2^-$ & 0 & 6.36 
             & 6.98 & 5.43E-4 & -1.61E-2 & 7.21E-2 & 5.24E-3 & 6.40 & 2.44E-3 & -8.03E-4 & 2.80E-2 & 1.75E-4 \\
  $^{92}$Rb  & 8.10 & 0    & $0^- \rightarrow 0^+$     & 0 & 5.75 
             & 7.63 & 1.40E-4 & 0 & 1.34E-1 & 0 & 6.71 & 1.20E-3 & 0 & 3.04E-2 & 0 \\
  $^{93}$Rb  & 7.47 & 0    & $5/2^- \rightarrow 5/2^+$ & 0 & 6.14 
             & 7.14 & 3.44E-4 & -6.25E-2 & 2.61E-2 & 1.13E-2 & 6.87 & 8.26E-4 & -5.18E-3 & 2.76E-2 & 7.16E-4 \\
  $^{94}$Y   & 4.92 & 0.92 & $2^- \rightarrow 2^+$     & 0 & 7.18 
             & 7.69 & 1.74E-4 & -1.66E-1 & 1.60E-2 & 2.03E-2 & 7.49 & 2.08E-4 & -2.46E-2 & 2.45E-2 & 2.98E-3 \\
  $^{95}$Rb  & 9.23 & 0.68 & $5/2^- \rightarrow 5/2^+$ & 0 & 6.01 
             & 7.46 & 6.59E-5 & 1.54E-2 & -3.83E-3 & 2.37E-2 & 7.56 & 1.47E-4 & 1.25E-3 & 2.65E-2 & 1.55E-3 \\
  $^{95}$Sr  & 6.09 & 0    & $1/2^+ \rightarrow 1/2^-$ & 0 & 6.16 
             & 7.82 & 9.04E-5 & 5.09E-4 & 1.14E-1 & 9.39E-5 & 7.01 & 5.96E-4 & 4.71E-6 & 2.95E-2 & 3.22E-6 \\
  $^{96}$Y   & 7.11 & 0    & $0^- \rightarrow 0^+$     & 0 & 5.59 
             & 8.91 & 7.29E-6 & 0 & 1.54E-1 & 0 & 7.90 & 7.63E-5 & 0 & 3.15E-2 & 0 \\
  $^{97}$Y   & 6.82 & 0    & $1/2^- \rightarrow 1/2^+$ & 0 & 5.70 
             & 7.71 & 1.10E-4 & 7.41E-4 & 1.13E-1 & 9.65E-4 & 6.92 & 7.30E-4 & -1.05E-4 & 3.03E-2 & 3.47E-5 \\
  $^{98}$Y   & 8.99 & 0    & $0^- \rightarrow 0^+$     & 0 & 5.8  
             &11.06 & 5.20E-8 & 0 & 2.11E-1 & 0 & 9.83 & 9.02E-7 & 0 & 3.36E-2 & 0 \\
  $^{133}$Sn & 8.05 & 0    & $7/2^- \rightarrow 7/2^+$ & 0 & 5.48 
             & 7.58 & 1.77E-4 & -6.22E-2 & 1.11E-1 & 5.38E-3 & 6.47 & 2.08E-3 & -9.69E-4 & 3.75E-2 & 7.64E-5 \\
  $^{135}$Te & 6.05 & 0    & $7/2^- \rightarrow 7/2^+$ & 0 & 6.14 
             & 6.85 & 1.22E-3 & -1.21E-1 & 4.66E-2 & 1.00E-2 & 6.15 & 4.34E-3 & -6.13E-3 & 3.44E-2 & 5.09E-4 \\
  $^{135}$Sb & 8.04 & 0    & $7/2^+ \rightarrow 7/2^-$ & 0 & 5.82 
             & 5.79 & 4.73E-3 & -2.70E-3 & -5.13E-3 & 1.46E-2 & 5.43 & 2.20E-2 & -6.71E-4 & 3.58E-2 & 4.61E-4 \\
  $^{136m}$I & 6.88 & 1.89 & $6^- \rightarrow 6^+$     & 0 & 6.25 
             & 7.60 & 1.11E-4 & 2.72E-2 & 1.45E-1 & 5.51E-3 & 6.72 & 1.17E-3 & 3.44E-4 & 3.48E-2 & 8.95E-5 \\
  $^{136m}$I & 6.88 & 2.26 & $6^- \rightarrow 6^+$     & 0 & 6.83 
             & 7.28 & 2.46E-4 & -2.52E-2 & -5.74E-2 & 1.46E-2 & 7.50 & 1.88E-4 & -1.38E-2 & 1.46E-2  & 3.22E-3 \\
  $^{137}$I  & 6.03 & 0    & $7/2^+ \rightarrow 7/2^-$ & 0 & ---  
             & 8.00 & 3.72E-5 & 1.63E-2 & -1.50E-2 & 1.25E-2 & 7.65 & 1.35E-4 & -4.10E-4 & 3.25E-2 & 5.34E-4 \\
  $^{142}$Cs & 7.33 & 0    & $0^- \rightarrow 0^+$     & 0 & 5.59 
             &10.94 & 6.49E-8 & 0 & 5.83E-1 & 0 & 8.96 & 6.65E-6 & 0 & 4.04E-2 & 0 \\
  $^{86}$Br  & 7.63 & 0    & $1^- \rightarrow 0^+$     & 1 & 7.5  
             & 7.40 & 7.04E-4 & -1.45E-1 & 3.12E-2 & 6.75E-3 & 8.04 & 1.64E-4 & -1.38E-1 & 3.53E-2 & 5.97E-3 \\
  $^{86}$Br  & 7.63 & 1.57 & $1^- \rightarrow 2^+$     & 1 & 7.2  
             & 6.80 & 1.94E-3 & -1.81E-1 & 2.31E-2 & 1.40E-2 & 7.61 & 4.31E-4 & -1.85E-1 & 3.94E-2 & 1.13E-2   \\
  $^{87}$Se  & 7.47 & 0    & $3/2^+ \rightarrow 5/2^-$ & 1 & 6.1  
             & 7.24 & 4.40E-4 & -1.50E-1 & 1.45E-2 & 1.41E-2 & 8.10 & 8.48E-5 & -1.57E-1 & 2.40E-2 & 1.15E-2 \\
  $^{89}$Br  & 8.26 & 0    & $5/2^- \rightarrow 3/2^+$ & 1 & 6.5  
             & 9.11 & 1.21E-5 & -1.29E-1 & 2.40E-2 & 5.88E-3 & 9.78 & 2.61E-6 & -1.24E-1 & 2.72E-2 & 5.37E-3 \\
  $^{91}$Kr  & 6.77 & 0    & $5/2^+ \rightarrow 3/2^-$ & 1 & 6.69 
             & 7.71 & 2.32E-4 & -1.50E-1 & 1.01E-2 &1.00E-2  & 8.47 & 4.55E-5 & -1.51E-1 & 1.64E-2 & 9.25E-3 \\
  $^{95}$Rb  & 9.23 & 0.56 & $5/2^- \rightarrow 7/2^+$ & 1 & 6.03 
             & 7.01 & 2.03E-4 & 5.14E-2 & -1.36E-3 & 1.67E-2 & 7.73 & 5.12E-5 & 8.85E-3 & 1.15E-2 & 1.28E-2 \\
  $^{134m}$Sb& 8.52 & 1.69 & $7^- \rightarrow 6^+$     & 1 & 6.27 
             & 7.34 & 7.35E-5 & 1.96E-1 & -8.76E-2 & 2.34E-2 & 8.17 & 1.01E-5 & 2.15E-1 & -8.56E-2 & 2.52E-2 \\
  $^{134m}$Sb& 8.52 & 2.40 & $7^- \rightarrow 6^+$     & 1 & 5.95 
             & 7.11 & 5.66E-4 & -9.54E-2 & -3.72E-2 & 9.52E-3 & 7.89 & 1.17E-4 & -1.19E-1 & -2.55E-2 & 9.24E-3 \\
  $^{136}$Te & 5.12 & 0    & $0^+ \rightarrow 1^-$     & 1 &$>$6.7
             & 6.81 & 4.19E-4 & 1.26E-1 & -1.02E-1 & 1.86E-2 & 7.66 & 6.16E-5 & 1.06E-1 & -1.01E-1 & 1.85E-2 \\
  $^{138}$I  & 7.99 & 0    & $1^- \rightarrow 0^+$     & 1 & 8.83 
             & 7.75 & 6.30E-5 & 2.27E-3 & -5.59E-2 & 9.37E-3 & 8.58 & 1.11E-5 & -2.13E-2 & -4.95E-2 & 8.31E-3 \\
  $^{140}$Xe & 4.06 & 0.08 & $0^+ \rightarrow 1^-$     & 1 & 6.14 
             & 6.26 & 1.69E-3 & 1.62E-1 & -1.03E-1 & 1.55E-2 & 7.11 & 2.40E-4 & 1.54E-1 & -1.06E-1 & 1.59E-2 \\
  $^{140}$Cs & 6.22 & 0    & $1^- \rightarrow 0^+$     & 1 & 7.05 
             & 8.63 & 6.16E-6 & 1.17E-1 & -8.46E-2 & 1.30E-2 & 9.47 & 9.09E-7 & 1.05E-1 & -8.43E-2 & 1.28E-2 \\
  $^{143}$Cs & 6.26 & 0    & $3/2^+ \rightarrow 5/2^-$ & 1 & $\approx$ 5.7 
             & 6.69 & 6.35E-4 & 3.49E-2 & 3.62E-3 & 1.56E-2 & 7.46 & 1.24E-4 & -6.80E-3 & 1.62E-2 & 1.56E-2 \\
  $^{88}$Rb  & 5.31 & 0    & $2^- \rightarrow 0^+$     & 2 & 9.25 
             & 8.11 & 8.22E-5 & -1.77E-1 & 0 & 1.55E-2 & 8.87 & 1.45E-5 & -1.77E-1 & 0 & 1.55E-2 \\
  $^{94}$Y   & 4.92 & 0    & $2^- \rightarrow 0^+$     & 2 & 9.35 
             & 8.38 & 4.44E-5 & -1.90E-1 & 0 & 1.79E-2 & 9.13 & 7.83E-6 & -1.90E-1 & 0 & 1.79E-2 \\
  $^{95}$Rb  & 9.23 & 0    & $5/2^- \rightarrow 1/2^+$ & 2 & $\geqslant$10.2 
             & 9.58 & 2.81E-6 & -1.05E-1 & 0 & 5.52E-3 &10.33 & 4.95E-7 & -1.05E-1 & 0 & 5.52E-3 \\
  $^{139}$Xe & 5.06 & 0    & $3/2^- \rightarrow 7/2^+$ & 2 & 8.6 
             & 7.28 & 5.59E-4 & -1.85E-1 & 0 & 1.70E-2 & 8.03 & 9.85E-5 & -1.85E-1 & 0 & 1.70E-2 \\
  \bottomrule
\end{tabular*}
\end{table*}

As the first application of a new model for the calculation of first-forbidden transition, we choose the 35 dominant first-forbidden transitions that are expected to mitigate the observed reactor anti-neutrino spectral shoulder to a large extent \cite{Hayen_PRC_R_2019, Hayen_PRC_2019}. These transitions, as shown in Table \ref{tab1}, involve even-even, odd-mass and odd-odd nuclei from medium-heavy (such as $^{86}$Br) to heavier (such as $^{143}$Cs) ones, including those are nearly spherical (such as $^{133}$Sn) and largely deformed (such as $^{94-98}$Y). Besides, transitions from both ground state and isomers (such as $^{136m}$I and $^{134m}$Sb) are involved. Compared with the list of transitions in Ref. \cite{Hayen_PRC_2019}, the spin-parity of the final state for transition from $^{95}$Rb is predicted and set by the PSM as $5/2^+$, and the spin-parity of the ground state of $^{89}$Br is determined and revised as $5/2^-$. 

It is seen from Table \ref{tab1} that without the quenching factors, the calculated Log$f_0 t$ for $|\Delta J| = 0, 1$ non-unique transitions are systematically larger than the data, while the calculated Log$f_0 t$ for $|\Delta J| = 2$ unique transitions are systematically reduced compared with the data. This indicates that the non-unique (unique) transitions are underestimate (overestimate) when no quenching factors are introduced. By comparison, when the SM quenching factors in Eq. (\ref{eq.quench}) are adopted, all the unique transitions are improved and well described by the PSM that the difference between calculated $f_0 t$ and the data is within a factor of four. This is due to the fact that unique transitions depend only on tensor operator of rank two, i.e., the one involved in the $z$ matrix element of Eq. (\ref{eq.all_ME}d), and the corresponding quenching factor $f_q(z) = 0.42$ is adopted by the SM in Eq. (\ref{eq.quench}). The SM quenching factors also improved the Log$f_0 t$ for $|\Delta J| = 0$ non-unique transitions so that 13 of the 18 transitions are within one order of magnitude when compared with the data, while the Log$f_0 t$ for $|\Delta J| = 1$ non-unique transitions are not improved. This indicates that the quenching factors are important for first-forbidden transitions and that different quenching factors should probably be involved in different nuclear models (SM, PSM, QRPA etc.). The quenching problem for allowed GT transition of $\beta$ decay and for the double $\beta$ decay has drew much attention recently \cite{Javier2011PRL, LJWang_current_2018_Rapid, quenching_nature_phys_2019, Suhonen_2019_Phys_Rep}, and the quenching problem for (first-) forbidden transitions deserve further investigation.

\begin{figure*}
\begin{center}
  \includegraphics[width=0.85\textwidth]{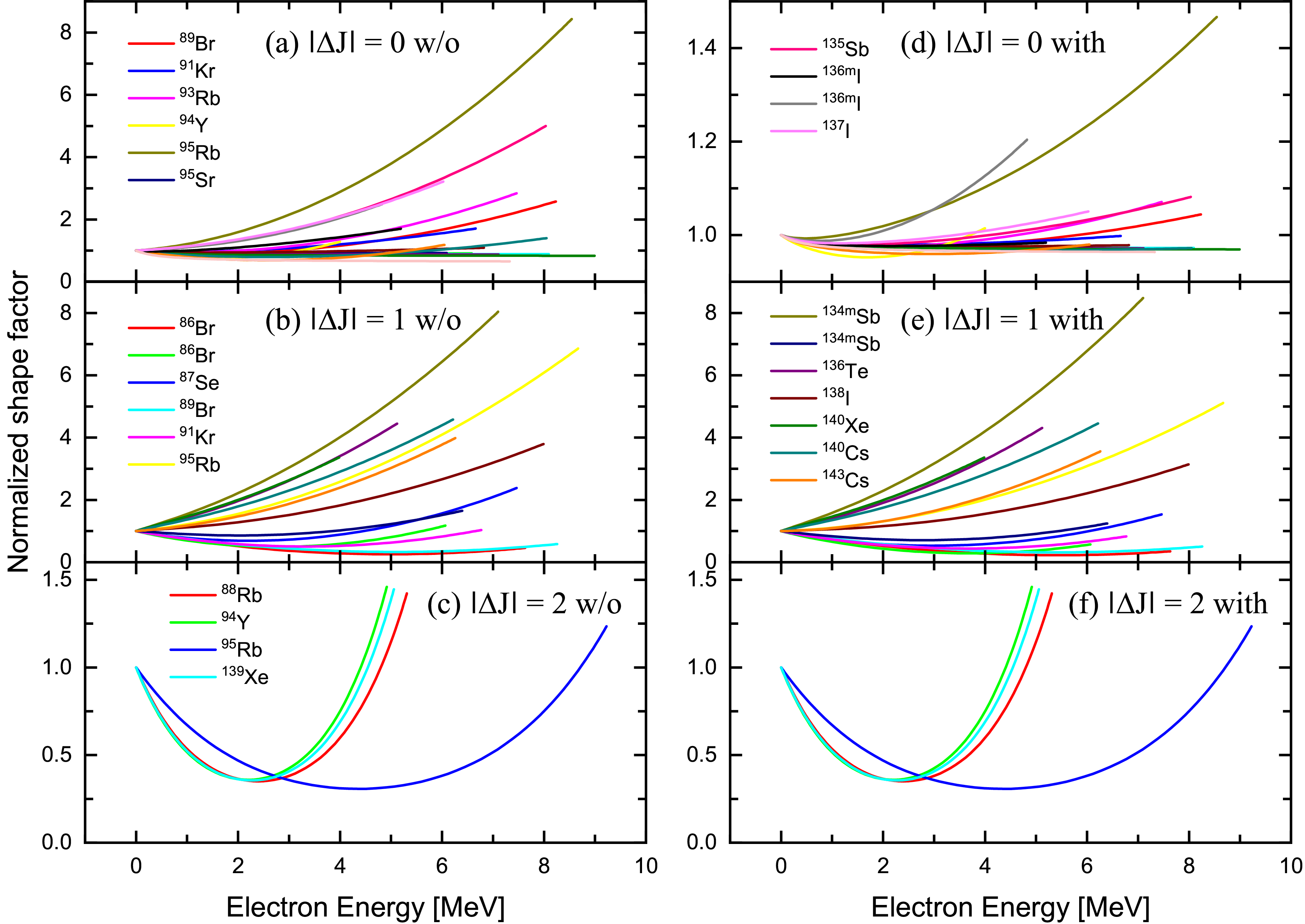}
  \caption{\label{fig:fig1} (color online) Calculated normalized shape factor for the transitions in Table \ref{tab1} as a function of the electron kinetic energy, by the PSM with and without the quenching factors. See the text for details. }
\end{center}
\end{figure*}

As discussed earlier, the reactor anti-neutrino spectral shoulder refers to the spectral disagreement (bump) between observations against theoretical predictions at $4-6$ MeV of anti-neutrino energy \cite{Ahn_PRL_2012, Abe_JHep_2014, An_PRL_2016}, where, it is noted that all forbidden transitions were either approximated as allowed (such as adopting $C(W)=1$) \cite{Huber_PRC_2011, Dwyer_PRL_2015} or assumed to have a unique shape \cite{Mueller_PRC_2011} in theoretical predictions and analyses. Recently, the shape factors of important first-forbidden transitions are calculated by the complicated SM method, the corresponding shape factors are found to deviate significantly from unity and to be different for different transitions, and the spectral shoulder problem is found to be mitigated \cite{Hayen_PRC_R_2019, Hayen_PRC_2019}. Therefore, first forbidden transitions are expected to be essential ingredient for better understanding the reactor anti-neutrino spectra and to merit additional research. 

In Table \ref{tab1} the coefficients $k, a, b, c$ in the shape factor of Eqs. (\ref{eq.CW}, \ref{eq.CW2}) from our PSM calculations are shown and the corresponding shape factors normalized to $C(W=1)=1$, i.e., $C(W)=1$ when the electron kinetic energy is zero, are illustrated in Fig. \ref{fig:fig1}. It is seen that the shape factors of most transitions deviate from unity and are different from each other. Besides, the quenching factors affect the shape factors to a large extent for non-unique transitions (note the different scales in the $y$ axis for $|\Delta J| =0$ case), but not for unique transitions as seen from Table \ref{tab1} and Fig. \ref{fig:fig1}(c) and (f). The reason for the latter is that unique transitions depend only on the $z$ matrix element of Eq. (\ref{eq.all_ME}d), then $k$, $ka$ and $kc$ coefficients depend only on $z^2$ terms as shown in Eq. (\ref{eq.kabc}) so that the $a, b, c$ coefficients do not depend on $z$ (and quenching factors). This is interesting as shape factors of these unique transitions are important for reducing the anti-neutrino spectra at low neutrino energy \cite{Hayen_PRC_R_2019, Hayen_PRC_2019}. Compared with the SM shape factors in Refs. \cite{Hayen_PRC_R_2019, Hayen_PRC_2019}, our PSM calculations give similar results for unique transitions while very different results for non-unique transitions. When quenching factors are adopted, the $^{95}$Rb and $^{136m}$I transitions for $|\Delta J|=0$ case and many transitions for $|\Delta J|=1$ case are shown to increase rapidly with the electron energy, which are different from the SM calculations in Refs. \cite{Hayen_PRC_R_2019, Hayen_PRC_2019}. The corresponding effect on our understanding of the reactor anti-neutrino spectral shoulder would be interesting and is planed as a future work.



In summary, we propose, as far as we know, a theoretical description of nuclear first-forbidden transitions based on angular momentum projection for the first time. The related model, the projected shell model, can now serve as an effective shell-model method for description of both allowed and first-forbidden transitions of nuclear $\beta$ decay for even-even, odd-mass and odd-odd nuclei from light to superheavy ones. As examples, 35 dominant first-forbidden transitions that are expected to be important for the reactor anti-neutrino spectra problems are calculated, where both the calculated Log$f_0 t$ values and related shape factors are found to be dependent on the quenching factors, indicating that the quenching factors for different nuclear matrix elements (transition operators) of first-forbidden transitions need further investigation as in the studies of allowed transitions in $\beta$ decay and of double $\beta$ decay. 

The model can be developed to calculate stellar weak-interaction rates of nuclei in high density and temperature environments, with contributions from both allowed and first-forbidden transitions, which is accomplished and will be published elsewhere soon.

\section*{Acknowledgements}
L.J.W. would like to thank G. Mart\'{\i}nez-Pinedo for valuable communications and Y. Sun for helpful discussions and encouragement about theoretical details, and B. S. Gao for discussions about experimental details, on forbidden transitions. This work is supported by the National Natural Science Foundation of China (Grants No. 12275225), by the Fundamental Research Funds for the Central Universities (Grant No. SWUKT22050), and partially supported by the Key Laboratory of Nuclear Data (China Institute of Atomic Energy).

\section*{Declaration of competing interest}
The authors declare that they have no known competing financial interests or personal relationships that could have appeared to influence the work reported in this paper.

\bibliographystyle{elsarticle-num}

\bibliography{first_forbidden}

\end{document}